\documentclass{SciPost}
 
\hypersetup{
    colorlinks,
    linkcolor={red!50!black},
    citecolor={blue!50!black},
    urlcolor={blue!80!black}
}
 
\usepackage[bitstream-charter]{mathdesign}
\DeclareSymbolFont{usualmathcal}{OMS}{cmsy}{m}{n}
\DeclareSymbolFontAlphabet{\mathcal}{usualmathcal}
 
\fancypagestyle{SPstyle}{
\fancyhf{}
\lhead{\colorbox{scipostblue}{\bf \color{white} ~SciPost Physics Codebases }}
\rhead{{\bf \color{scipostdeepblue} ~Submission }}

\fancyfoot[C]{\textbf{\thepage}}
}

\usepackage{hyperref}
\usepackage{orcidlink}
\usepackage{cleveref}
\usepackage{fontawesome}
\usepackage[frozencache=true, cachedir=minted-cache]{minted}
\AtBeginEnvironment{minted}{}
\usepackage{verbatim}             
\usepackage{fvextra}           
\usepackage{xcolor}           
\usepackage{adjustbox}       
\usepackage{graphicx}
\usepackage{caption}          
\usepackage{float}             
\usepackage{xparse}    
\usepackage{changepage}
\usepackage{subcaption}

\newcommand{\mintedsmall}{\fontsize{4pt}{4.8pt}\selectfont}
 
\definecolor{bg}{rgb}{0.95, 0.95, 0.95} 
 
\setminted{
    linenos,
    frame=lines,
    framesep=1mm,
    baselinestretch=1.1,
    bgcolor=bg,
    fontsize=\footnotesize,
    breaklines,
    breakanywhere,
}
 
\newcommand{\python}[1]{%
  \mintinline[breaklines, breakanywhere]{python}{#1}%
}
 
\newenvironment{pythonTable}
  {\VerbatimEnvironment
   \begin{minipage}[t]{\linewidth}%
   \begin{minted}[fontsize=\footnotesize, breaklines, breakanywhere, frame=none,linenos=false]{python}}
  {\end{minted}%
   \end{minipage}}
 
\begin{document}
 
\pagestyle{SPstyle}
 
\begin{center}{\Large \textbf{\color{scipostdeepblue}{
Wigner Time: from experimental intent to real-time laboratory control
\\
}}}\end{center}
 
\begin{center}\textbf{
T.~W.~Clark\textsuperscript{1$\star$}\orcidlink{0009-0008-8229-6242},
B.~Sárközi\textsuperscript{2},
A.~Dombi\textsuperscript{1},
Á.~Kurkó\textsuperscript{1},
D.~Nagy\textsuperscript{1},
A.~Simon\textsuperscript{1,3},
D.~Varga\textsuperscript{1,3},
P.~Domokos\textsuperscript{1,4},
A.~Vukics\textsuperscript{1$\dagger$}\orcidlink{0000-0001-8916-4033} 
}\end{center}
 
\begin{center}
\textbf{1} HUN-REN Wigner Research Centre for Physics, H-1525 Budapest, PO Box 49, Hungary\\\textbf{2} Center for Hybrid Quantum Networks (Hy-Q), Niels Bohr Institute, University of Copenhagen, Jagtvej 155A, Copenhagen DK-2200, Denmark\\\textbf{3} Department of Physics of Complex Systems, ELTE Eötvös Loránd University, Pázmány Péter sétány 1/A, Budapest 1117, Hungary\\\textbf{4} Department of Theoretical Physics, Institute of Physics, Budapest University of Technology and Economics, Műegyetem rkp. 3, Budapest 1111, Hungary
\\[\baselineskip]
$\star$ \href{mailto:thomas.clark@wigner.hun-ren.hu}{\small thomas.clark@wigner.hun-ren.hu} $\dagger$ \href{mailto:vukics.andras@wigner.hun-ren.hu}{\small vukics.andras@wigner.hun-ren.hu}
\end{center}

\section*{\color{scipostdeepblue}{Abstract}}
\textbf{\boldmath{%
Precisely timed, multi-device control in atomic, molecular and optical physics is
provided by specialized real-time systems, which impose their own terms on the
description of the experiment: a program over a global clock, in which every stage
boundary is an absolute time computed from everything preceding it. Such
descriptions are tightly coupled – changing one stage affects all later ones –
and are correspondingly hard to reuse, inspect, or move elsewhere. We introduce
Wigner Time, a Python package in which the experimental procedure is instead
represented as data: a table of timed updates, assembled by composing functions and
referred to named points in the experiment rather than to absolute instants.
Hardware enters only at a final conversion step, leaving the description readable
and back-end agnostic. Wigner Time has run two cold-atom setups for more than two
years, where taking the computation out of the real-time program improved the
achievable temporal resolution five-fold; the design applies to any domain requiring precise
multi-device timing.
}}
 
\vspace{\baselineskip}


 
\vspace{10pt}
\noindent\rule{\textwidth}{1pt}
\tableofcontents
\noindent\rule{\textwidth}{1pt}
\vspace{10pt}

 
\section{Introduction}
\label{sec:intro}
Experimental setups in atomic, molecular, and optical (AMO) physics are now the most precisely controllable systems ever produced by mankind \cite{PhysRevLett.132.190001,Bluvstein2024}. Accordingly, these systems require a high level of electronic control, where timed signals, both analog and digital, command a diverse array of components, such as laser frequencies, optical shutters, acousto-optical modulators, electromagnets and optomechanical stages. In this context, however, “precisely” routinely means microsecond repeatability or, in the most demanding experiments, even sub-microsecond resolution – ruling out general-purpose workstations. Instead, specialized \emph{real-time} I/O systems are necessary, either in the form of PCI measurement cards, custom FPGA-based modules, or self-standing units.
 
By eschewing general-purpose machines and even linear programming languages,
however, such systems impose their own terms on the description of the experiment.
Typically the only available notion of time is a global cycle counter, so every
stage boundary must be computed as a running sum of all preceding durations, and
the body of the experiment becomes a flat dispatch on that counter – with the
finite response times of real hardware, shutters in particular, folded into those
sums by hand. Each such expression is reasonable in isolation; together they
constitute a single, globally coupled arithmetic problem, in which the instants
that matter physically appear nowhere in the code, existing only as intermediate
values. \Cref{fig:before_after} gives a representative example, taken from the
implementation that the present work replaced.
 
\begin{figure}
\centering
\begin{minipage}{.9\linewidth}
\textbf{(a) ADbasic: the optical-pumping stage as previously implemented}
\begin{minted}[fontsize=\mintedsmall, frame=lines, breaklines, breakanywhere, linenos=false]{basic}
'=== defs.inc =========================================================
#Define OP_ExposureTime      FPar_17   ' in us
#Define OP_Shutter1OnDelay   1.5       ' in ms
#Define OP_Shutter2OnDelay   2.8       ' in ms
#Define ShutterResponseTime  4         ' in ms
#Define OP_AOM_Channel       30
#Define OP_Shutter1Channel   14
#Define OP_Shutter2Channel   15
 
'=== MOT_full.bas, file scope =========================================
#Define OP_Shutter2OffDelay (ShutterResponseTime-OP_Shutter2OnDelay)
 
'=== MOT_full.bas, LOWINIT ============================================
endOfMOT_Growth = secToCC(MOT_GrowTime)
endOfMolasses   = endOfMOT_Growth + msToCC(PolgradCoolingTime)
endOfOP         = endOfMolasses   + usToCC(OP_ExposureTime)
 
'=== MOT_full.bas, EVENT: dispatch on the global cycle counter =========
If (CycleCount = endOfMolasses-msToCC(SafetyMargin*OP_Shutter1OnDelay)) Then
  OP_Shutter1(1)
EndIf
 
If (CycleCount = endOfMolasses-usToCC(RampCoilsToDipoleInterval)) Then
  startCoilsRamp(usToCC(RampCoilsToDipoleInterval), HomogeneousMagneticFieldCurrent, -HomogeneousMagneticFieldCurrent, 0., 0.)
EndIf
 
If (CycleCount = endOfOP-msToCC((2.-SafetyMargin)*OP_Shutter2OffDelay)) Then
  OP_Shutter2(0)
EndIf
 
If (CycleCount = endOfOP) Then
  OP_AOM(0)
  repumpAOM(0) : repumpShutter(0)
  startCoilsRamp(usToCC(RampCoilsBackToQuadrupole), ...)
EndIf
 
'=== MOT_full.bas, EVENT, some 150 lines further down =================
If ((CycleCount >= rampCoilsBeginCC) And (CycleCount <= rampCoilsEndCC)) Then
  rampCoilsTemp = Tanh(TanhInterval*(2.*(CycleCount-rampCoilsBeginCC)/(rampCoilsEndCC-rampCoilsBeginCC)-1.))
    /TwoTimesTanhOfTheTanhInterval
  ...
EndIf
 
'=== tools.inc ========================================================
Sub startCoilsRamp(interval,lower,upper,lowerP,upperP)
  If (rampCoilsEndCC>CycleCount) Then error("overlapping coils ramps")
  rampCoilsBeginCC = CycleCount
  rampCoilsEndCC   = rampCoilsBeginCC + interval
  ...
EndSub
\end{minted}
\end{minipage}
 
\vspace{\baselineskip}
 
\begin{minipage}{.9\linewidth}
\textbf{(b) Wigner Time: the same stage}
\begin{minted}[fontsize=\mintedsmall, frame=lines, breaklines, breakanywhere, linenos=false]{python}
def optical_pumping(  # assumes the repump is on
    duration_exposition=80e-6,
    duration_coil_ramp=50e-6,
    i=-0.12,
    delay1=0, delay2=0, delay_repump=0,
    delay_shutter_reinitialization=0.1,
):
    duration_full = duration_exposition + duration_coil_ramp
    return tl.stack(
        tl.ramp(coil__MOT_lower__A=i, coil__MOT_upper__A=-i,
                duration=duration_coil_ramp),
        tl.update(AOM__OP=[[-0.1, 0], [duration_coil_ramp, 1], [duration_full, 0]]),
        tl.update(shutter__OP1=[
            [duration_coil_ramp - constants.OP.lag_shutter_on + delay1, 1],
            [delay_shutter_reinitialization, 0]]),
        tl.update(shutter__OP2=[
            [duration_full - constants.OP.lag_shutter_off + delay2, 0],
            [delay_shutter_reinitialization, 1]]),
        tl.update(shutter__repump=0,
                  time=duration_full - constants.lag_repump_shutter + delay_repump),
        tl.update(AOM__repump=0, time=duration_full),
        tl.anchor(duration_full),
        context="optical_pumping",
    )
\end{minted}
\end{minipage}
 
\caption{\label{fig:before_after} One experimental stage, before and after. In
(a), the stage is distributed across five locations: hardware channels and
shutter response times in an include file, a derived delay at file scope, its
temporal boundaries as running sums in \texttt{LOWINIT}, its actual events as
equality tests against a global cycle counter in \texttt{EVENT}, and the
interpolation of its coil ramp in a separate block far below, reached through
reserved global state set by a subroutine in a third file. Every instant is
absolute, so lengthening any earlier stage silently displaces this one; units
survive only in comments, and the choice between \texttt{msToCC} and
\texttt{usToCC} is unchecked; and because the ramp occupies a single set of
globals, a second concurrent coil ramp is a runtime error rather than a new
operation.
In (b), the same stage is one function. Times are relative to its own
beginning, values carry physical units in the variable names, the shutter lags
are named constants offset by explicitly exposed drift-compensation parameters,
and the closing \python{anchor} means the stage can be inserted anywhere in any
timeline without any of its internal arithmetic changing. Here, as in the
appendices, \python{tl} stands for the module \python{wignertime.timeline}.
}
\end{figure}

\begin{figure}
  \centering
  \includegraphics[width=.95\linewidth]{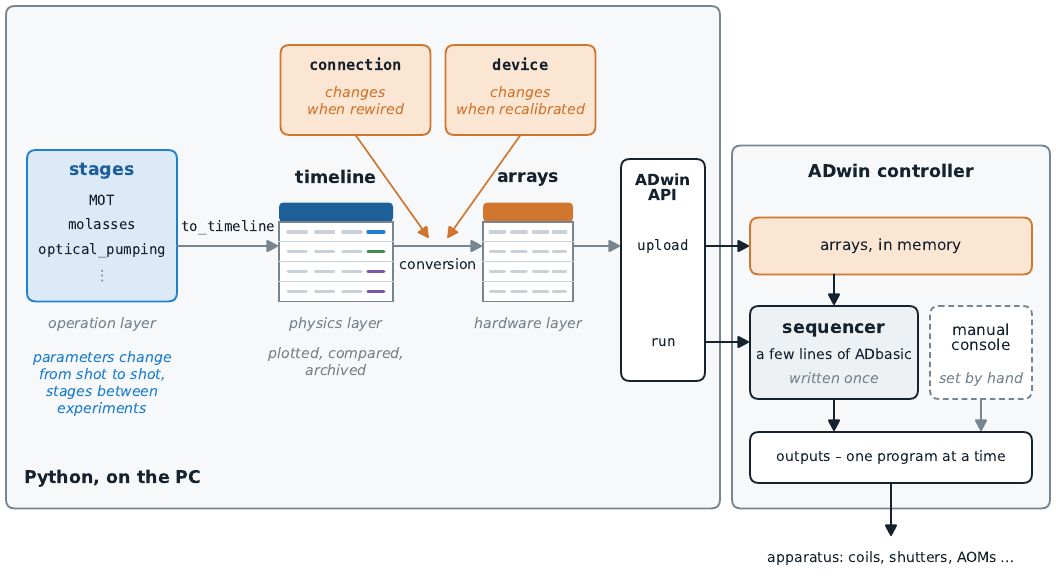}
  \caption{\label{fig:hardware_overview} How Wigner Time connects Python on the PC to the ADwin controller. The experiment is described entirely on the PC, in the three layers of \cref{sec:definitions}. Stages compose a timeline in physical units, which can be plotted, compared and archived like any other table, and the conversion step combines it with the \python{connection} and \python{device} tables into hardware-ready arrays. This is the only point at which hardware detail enters. Through the ADwin API, \python{upload} writes the arrays into the controller’s memory, and \python{run} starts the sequencer, which plays them. The manual console shares the outputs with the sequencer, one program at a time (\cref{sec:adwin_operation}). The colors mark how often each part is rewritten: the parameters of the stages change from shot to shot, and the stages themselves between experiments; the \python{connection} table changes when the apparatus is rewired, and the \python{device} table when a device is recalibrated; the sequencer is written once.}
\end{figure}

Wigner Time bridges the gap between user-friendly design of experimental procedures and the requirements of a precise hardware timing system, e.g. ADwin (\cref{fig:hardware_overview}). Above all, it is a Python package for high-level representation and manipulation of such procedures, following a data-oriented approach that has recently been gaining prominence, particularly in synergy with the more general resurgence of functional programming \cite{dop2022, hanson2021software, hickey2008clojure}. Wigner Time therefore does not define any new object types, but is rather based around the idea of a \emph{timeline}: at heart, simply a table of rows and columns. The rows represent operations or events, and are characterized by the parameters represented by the columns, with four core headings: \python{variable} (the name of a quantity or hardware channel), \python{time}, \python{value} and \python{context} (a concise description of a real-world situation, e.g. an experimental stage, like \python{"optical_pumping"}). The timeline grows naturally by adding new rows.
 
By grounding the framework in a foundational table structure, we can benefit from decades of database development, particularly in-memory database-like systems. When in doubt, therefore, the user can simply manipulate their timeline using the well-developed pandas ecosystem completely independently of our package. For most operations, however, this will not be necessary, as Wigner Time provides layers of convenience functions.
 
Existing solutions to the control problem fall broadly into two families. The first co-designs software together with its own hardware. An early and influential example is the laboratory control system of Meyrath and Schreck \cite{meyrath_schreck_control}, which distributes data from a central computer over a parallel bus to purpose-built analog, digital and radio-frequency output boards, with circuit designs and software released for others to rebuild. The same philosophy, at greater scale, characterizes the Sinara hardware ecosystem and its ARTIQ control system \cite{kasprowicz2020artiq,bourdeauducq2016artiq}: a modular open-hardware platform organized around an FPGA carrier, programmed in a Python-derived language that is compiled and executed on the hardware itself. The second family provides software over commercial off-the-shelf hardware, and divides in turn by how the experiment is described. Cicero/Atticus \cite{keshet2013distributed} uses a graphical editor, in which a sequence is composed of successive “words” – each beginning when the previous one ends – and dispatched through a client–server split to National Instruments (NI) output cards. labscript \cite{starkey2013labscript,labscript} and the Entangleware Sequencer \cite{kowalski2023entangleware} are instead scripted: experiments are written as Python programs that emit low-level hardware instructions, in the former case timed by a pseudoclock across a range of supported vendors, in the latter compiled to a bitstream for an NI FPGA. Alongside all of these sit the private, lab-specific programs commonplace in cold-atom groups, of which little is ever published.
 
Wigner Time cuts across this division, because the distinction that matters to us is not which hardware is targeted but what the experimental description \emph{is}. In every system above, the description is a \emph{program}: an imperative script, a graphical program, or a compiled real-time kernel, which is then executed in order to \emph{emit} hardware instructions. In Wigner Time the description is \emph{data} – a table of rows – and the library is a set of functions that transform tables. Hardware enters at exactly one point, the conversion step of \cref{sec:hardwarelayer}. This is the origin of the advantages listed below: portability and inspectability follow from the timeline being a value rather than a side effect, and hardware detail – module numbers, channel assignments, conversions to DAC codes – no longer has to sit next to the physics. The properties below each answer some part of this: the global coupling of times, the mixing of hardware detail with the physics, and the fact that a program, once run, leaves nothing behind to inspect.

\paragraph{Easy} In Hickey’s now-standard distinction \cite{Hickey2011Simple},
a tool is \emph{easy} if it is familiar and near at hand. Specialized timing
systems fail this test by construction: they result in specialized languages,
which bear two contrasting characteristics of low-level programming languages: power and precision versus diminished intuitive clarity. Wigner Time, by
contrast, lets the user design timelines in a general-purpose, popular language,
with the general-purpose, popular pandas ecosystem underneath.
 
\paragraph{Simple} A tool is \emph{simple}, in the complementary sense, if its
concerns are not intertwined – and easy tools often fail here. Wigner Time is
built around one plain data structure, with each concern in its own place:
physical calibration (\python{device}) is independent of wiring
(\python{connection}); timeline design is independent of hardware conversion;
ramps are stored symbolically and only expanded at the boundary. Every
convenience is an optional layer over the table, so one can always drop down an
abstraction level.
 
\paragraph{Portable} The inherent \emph{simplicity} means that the essential data is always accessible and transferable to any other language or collaborator. This also extends to integration with other scientific Python packages, the ubiquitous Jupyter system in particular. Ultimately, a single Jupyter notebook can be used to define complete workflows: from batches of experimental runs through data collection and analysis to publication-quality figures. By implementing the system on top of the most widely used technologies the core is well-tested and applicable to a wide range of problems. Furthermore, the table format makes it easy and efficient to store timelines.
 
\paragraph{Fast} Decoupling the timeline design from the implementation also means
that the real-time system program (\cref{fig:hardware_overview}), as distinct from
the Python code, is maximally simple. As almost all of the logic happens outside
the hardware, the low-level code can be reduced to a single concept: “at this cycle,
write the values listed for it.” Which values need writing at all – those that
differ from what an output already holds – is decided beforehand, in Python. This matters because every instruction in the event loop must
complete within one cycle, so logic in the real-time program is paid for directly
in temporal resolution. The effect is concrete: the ADbasic implementation this
work replaced evaluated ramp trajectories – including transcendental functions –
inside its event loop, alongside several dozen conditional tests, and ran at a
cycle time of 5\,µs; the present system runs the same experiments at 1\,µs. The
gain is not free, but rather a deliberate exchange of computation for memory,
since ramps expanded ahead of time occupy rows that were previously recomputed on
every cycle. On current hardware this is a favorable trade by a wide margin.
These are the inherent speed gains, but the development time saved by programming
in a high-level language like Python is an equally significant advantage.
 
\paragraph{Flexible} By combining a functional, data-oriented and bottom-up programming approach, Wigner Time makes it straightforward to add and modify entries in the timeline, responding naturally to rapidly evolving lab requirements. The resulting timeline can also serve purposes beyond running experiments – notably visualization and interrogative debugging, with basic display conveniences included in the package.
 
\paragraph{Curatable} Experimental stages are ordinary functions, so the quantities one expects to vary are their arguments – declared once, with defaults, alongside the code they govern – while everything that rarely changes is enclosed in the body. A stage’s interface is therefore self-documenting and discoverable by the usual Python means, and a specification needed in more than one place, such as a safe default state for the apparatus, can be written once and reused. Because the composition and its parameters live together in ordinary source, a complete experiment becomes a single versionable artifact: the difference between two campaigns is a diff, and the timeline actually delivered to the hardware can be archived alongside the data it produced.
 
\bigskip
With these goals in mind, the package has been under development at the Wigner Research Centre for Physics, Budapest, and so was named in honor of the Hungarian-born Nobel laureate: \emph{Eugene Wigner}. Having been in use for more than two years across two cold-atom setups, we now present it as an open-source package, distributed under the GNU General Public License v3 \cite{github}. In the rest of this paper, we first describe the high-level Python API in a back-end agnostic manner (\cref{sec:pythonAPI}) before presenting the conversion features specific to the ADwin timing system (\cref{sec:adwin}). We conclude with an overview and future outlook, before exhibiting concrete cold-atom laboratory examples in the appendices.
 
\section{Python API}
\label{sec:pythonAPI}
The following description relates to the current implementation, which uses \python{pandas} and assumes ADwin as the timing target. While the raw output can already be used with non-ADwin systems (e.g. National Instruments), these are not yet explicitly supported or optimized. For up-to-date API documentation, consult the automatically generated help files \cite{wignertime_docs}.
 
\begin{figure}
  \centering
  \includegraphics[width=.9\linewidth]{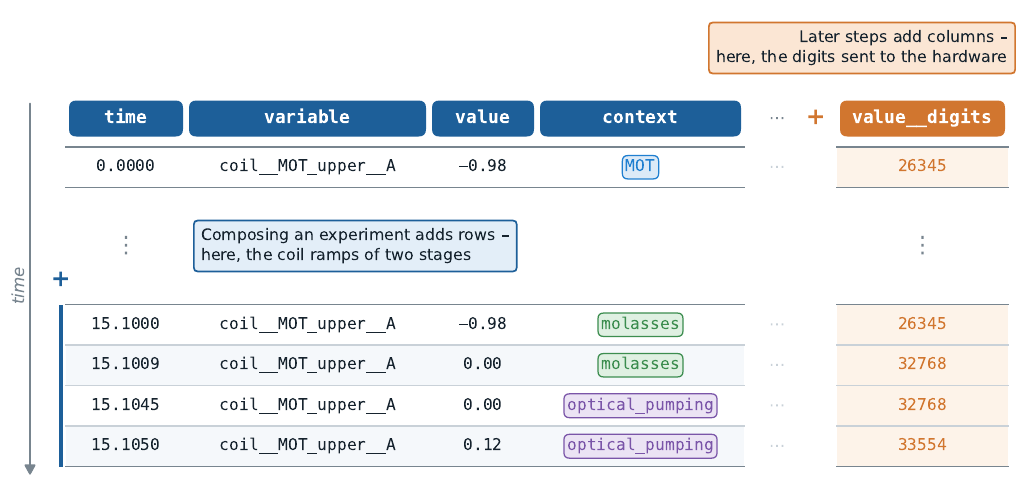}
  \caption{\label{fig:timeline_overview} The anatomy of a timeline. Each row records a single update of a single variable at a single instant, characterized by the parameters held in the columns. Composing an experiment adds rows. Columns can also be added, but this is rare in ordinary use and is normally done by the package itself – the conversion step, for instance, appends the digitized values sent to the hardware.}
\end{figure}

\subsection{Overview and definitions}
\label{sec:definitions}
 
Three layers of abstraction are easily identifiable in a typical Wigner Time workflow. The operation layer is client code that the package facilitates but does not prescribe. The physics and hardware layers are provided by the package itself. These might look like
\begin{description}
    \item[operation] “take a fluorescence image”
    \item[physics] “set laser 1 to 5\,W”
    \item[hardware] “send 5\,V to connection 2”
\end{description}
Timelines sit at the heart of every layer. They are the output of the operational layer (\cref{sec:operationlayer}), the core data structure of the physics layer (\cref{sec:physicslayer}), and the input to the hardware layer, where they are transformed into hardware-ready arrays (\cref{sec:hardwarelayer}).

As an example, consider controlling a cooling laser with an optical shutter, an acousto-optical modulator (AOM), and a beat lock with variable offset connected to the real-time system. The relevant functions will be introduced as needed, with a full description following in subsequent sections.
 
The first step is to define the experimental system. For digital channels, this means \emph{naming} the physical connections – associating module and channel numbers with their intended purpose. For analog channels, the same applies, with the addition of a unit. This distinguishes analog from digital channels, as required by most timing systems.
\begin{minted}[fontsize=\footnotesize, breaklines, frame=none]{python}
from wignertime.adwin import connection as adcon
from wignertime import device
from wignertime import conversion as conv
 
connections = adcon.new(
    ["shutter__MOT", 1, 11],
    ["AOM__MOT", 1, 1],
    ["AOM__MOT__transmission", 3, 1],
    ["lockbox__MOT__MHz", 3, 8],
    )
\end{minted}
Although the form of these names can be arbitrary, the package by default assumes and enforces the format \texttt{<device>\_\_<UID>(\_\_<unit>)}, where \texttt{<...>} denotes a user-defined identifier and \texttt{(...)} indicates an optional component. The device and the unit are single words, while the UID, a unique identifier, may contain single underscores, as in \python{coil__MOT_lower__A}. Any alternative convention should be applied consistently.
 
For analog connections, conversions and value ranges must also be specified. The conversion can be a linear factor, a function, or an interpolation from a calibration file – in each case mapping the conceptual unit to a voltage, which we assume to be the unit of the actual real-time controller analog outputs. The remaining arguments define the minimum and maximum permitted values, which are enforced before output. Omitted bounds default to $\pm\infty$.
\begin{minted}[fontsize=\footnotesize, breaklines, frame=none]{python}
devices = device.new(
    ["lockbox__MOT__MHz", 0.05, -200, 200], 
    # 0.05 V/MHz: +/-200 MHz spans +/-10 V
    [
        "AOM__MOT__transmission",
        conv.function_from_file(
            "resources/calibration/aom_calibration.dat",
            sep=r"\s+",
        ),
        0.0,
        1.0,
    ])
\end{minted}
 
Next, it is useful to specify the aspects of the experiment that change rarely, such as the initial and final device states. This can be done concisely, e.g.
\begin{minted}[fontsize=\footnotesize, breaklines, frame=none]{python}
import math
from wignertime.timeline import *

def default_state(**kwargs):
    return update(
        shutter__MOT=0,  # closed
        AOM__MOT=1,  # kept on, in thermal equilibrium
        lockbox__MOT__MHz=0.0,
        **kwargs,
    )

initial = default_state(time=-math.inf, context="ADwin_LowInit")
final = default_state(time=math.inf, context="ADwin_Finish")
\end{minted}
Here, \python{update} allows the user to specify common arguments such as the \python{time} and the \python{context}, as well as experiment-specific variables like \python{shutter__MOT} – all with a high signal-to-noise ratio in the syntax. The initial and final states are the same, so they are written once and placed twice. They belong to no instant of the experiment, and are therefore placed before and after it, at $-\infty$ and $+\infty$ (\cref{sec:context}).

The key experimental procedures are then defined:
\begin{minted}[fontsize=\footnotesize, breaklines, frame=none]{python}
MOT = update(
    shutter__MOT=1,
    context="MOT",
)
detuned_growth = ramp(
    lockbox__MOT__MHz=-5,
    duration=10e-3,
)
\end{minted}
None of these is yet a timeline. Each is a \emph{stage}: a description of what happens, written relative to its own beginning, which can later be placed anywhere. Stages are composed into a complete experiment in a readable and modular fashion, and \python{to_timeline} makes a timeline of the result:
\begin{minted}[fontsize=\footnotesize, breaklines, frame=none]{python}
tline = to_timeline(
    stack(
        initial,
        MOT,
        detuned_growth,
        final,
    )
)
\end{minted}
By design, each component automatically follows the end of the previous one, forming a causal chain through the \python{origin} mechanism (see \cref{sec:origin}). The result is a \python{pandas.DataFrame}, which can be manipulated directly when convenient: \python{tline[tline["context"] == "MOT"]}, for instance, selects the rows of the stage that forms a magneto-optical trap (MOT). The timeline can then be converted to an ADwin-compatible format and uploaded with \python{wignertime.adwin.core.upload} (\cref{sec:adwin}).

\subsubsection{The operation layer}
\label{sec:operationlayer}
The operation layer is where the experiment is described in its own terms – as a sequence of \emph{stages}, written as custom functions. For a cold-atom setup, these might include initialization, magneto-optical trapping, transport, imaging, and finalization.

This is the layer where most day-to-day experimental work takes place. In the example above, it is represented by the simplified MOT and detuned growth procedures, each of which would typically be wrapped in a more comprehensive function covering additional parameters.
 
The operation layer is not part of Wigner Time itself, but is rather client code – such as the examples in \cref{sec:demonstration} – that the package is designed to facilitate. The main recommendation is to define the building blocks of the experiment as a series of functions returning a stage, which can then be composed using \python{stack} (\cref{sec:stacking}).

That this layer is worth building is borne out elsewhere: in the CIRCUS system \cite{volponi2024circus}, an experiment class and accompanying function libraries were written on top of ARTIQ/Sinara precisely in order to reduce run scripts to a few readable lines.

\subsubsection{The physics layer}
\label{sec:physicslayer}
The physics layer is the core abstraction of Wigner Time. It is embodied by the variable-time-value-context (vtvc) \python{DataFrame} – the \emph{timeline} – where all values are expressed in real physical units (MHz, Amperes, etc.). Each row represents a single state change (\emph{update}) of a single variable at a given time. Variables may be physical quantities controllable via the real-time system, or virtual variables such as anchors (\cref{sec:anchor}) that exist only as reference points.
 
A fully assembled timeline represents a single experimental run (or “shot”, in the language of labscript \cite{starkey2013labscript}), storing a time-tagged list of updates. Currently implemented as a \python{pandas.DataFrame}, it is well suited to creating and extending experimental procedures (\cref{fig:timeline_overview}). It has at least the following columns:
\begin{description}
\item[variable : \python{str}] Names a single degree of freedom of an experimental apparatus – usually a physical variable controllable via the real-time system, e.g. \python{AOM__probe__W} or \python{AOM__probe}. Following the naming convention \texttt{<device>\_\_<UID>(\_\_<unit>)}, the presence or absence of a unit suffix distinguishes analog from digital variables.
\item[time : \python{float}] The time instant at which the given variable is to be updated. This can be negative when using \python{origin}s (\cref{sec:origin}). The initial and final states of the apparatus, which belong to no instant, are at $-\infty$ and $+\infty$ (\cref{sec:context}).
\item[value : \python{float}] The value to which the given variable is to be updated. Digital variables accept only 0 or 1; other values will be rounded, and incompatible types will be coerced or raise an error.
\item[context : \python{str}] A label for an experimental stage, e.g. \python{"magnetic_trapping"} or \python{"imaging"}, aiding readability and enabling context-based time references.
\end{description}
 
\subsubsection{The hardware layer}
\label{sec:hardwarelayer}
The hardware layer is produced by the ADwin-specific conversion function, which transforms the physics-layer timeline into arrays of numbers suitable for direct upload to the real-time controller.

The conversion translates each row one part at a time, drawing on the two tables defined above. In “set laser 1 to 5\,W”, the \python{connection} table translates the name: laser 1 is wired to connection 2, a module and channel number of the controller. The \python{device} table translates the value: it holds, for each analog variable, the conversion from physical units to voltage and the permitted range, so 5\,W becomes 5\,V, which is then digitized for the module. The time becomes a count of the controller’s cycles. The two tables are kept apart, so that recalibrating a device and rewiring the apparatus are independent operations, each requiring changes in only one place. Only the \python{connection} table is specific to ADwin.

This separation means that the timeline – expressed in physical units and real times – remains human-readable and backend-agnostic up to the point of conversion. Adapting Wigner Time to a different real-time controller requires a new conversion function at this final step, together with a program on the controller that consumes the resulting table. In the ADwin case the latter is the few lines of \cref{sec:adwin}; for other systems it may be a more substantial undertaking, depending on how directly the hardware can be driven from a precompiled list of updates.
 
\subsection{Origin}
\label{sec:origin}
Timelines are most naturally constructed from relative rather than absolute references – a stage should begin when the preceding one ends, without its absolute instant being known in advance. Every core function therefore accepts an origin keyword, which names the point in the existing timeline against which the new entries are measured. The mechanism is deliberately flexible, and is specified in full in \cref{sec:origin_full}; in day-to-day use, however, the keyword rarely appears in user code at all, because two cases cover almost everything.
 
\paragraph{The default} With the default origin, the reference point is the most recent \python{anchor} (\cref{sec:anchor}) if the timeline contains one, the most recent entry otherwise, and – failing both, as at the start of an experiment – absolute zero. Since every stage is recommended to end with an anchor, the practical consequence is that each \python{time} argument is read as a $\Delta t$ from the end of the preceding stage. This one default is what allows stages to be written purely in terms of their own internal logic and composed afterwards (\cref{sec:stacking}): no arithmetic inside a stage refers to anything outside it, and lengthening or inserting an earlier stage displaces everything that follows automatically, rather than invalidating a chain of hand-computed sums.
 
\paragraph{A context as origin} Passing the name of a \python{context} (\cref{sec:context}), e.g. \python{origin="molasses"}, measures the new entries from that stage rather than from the immediately preceding one – specifically from its anchor if it has one, and from its last entry otherwise. This is the mechanism behind interweaving (\cref{sec:interweaving}): a diagnostic, imaging or trigger operation can be attached to a named point of an existing timeline without restructuring the stages around it, and without the offset being worked out by hand.
 
The \python{origin} mechanism is deliberately an \emph{easy} layer over a \emph{simple} substrate: its flexibility exists only at construction time and never leaks into the data. Once resolved, the timeline contains nothing but absolute times and values, and any origin specification could equally have been written as an explicit numeric coordinate. The generality catalogued in \cref{sec:origin_full} is thus transient and optional, rather than a permanent property of the resulting artifact.
 
\subsection{Core functions}
\label{sec:functions}

The three main convenience functions are \python{update}, which sets variables at given instants; \python{ramp}, which transitions variables to new values over time; and \python{anchor}, which places the virtual reference rows that the \python{origin} mechanism of \cref{sec:origin} resolves against. Each of them returns a \emph{stage}, and \python{to_timeline} makes a timeline of a stage (\cref{sec:to_timeline}).

\subsubsection{Update}
\begin{minted}[fontsize=\footnotesize, breaklines, frame=none, linenos=false]{python}
def update( *, time=0.0, context=INFER, origin=INFER, **vtvc_dict )
\end{minted}
This function accepts a wide range of input formats through the \python{**vtvc_dict} keyword argument, whose possibilities are laid out in \cref{tab:inputSpecs}. The arguments \python{time} and \python{context} serve as defaults for cases where time and context are not specified individually for each variable. Any other keyword is read as the name of a variable – a property put to use in \cref{sec:forwarding}. The asterisk in the signature means that the arguments after it can be given only by name, so that a value cannot end up in the wrong argument because of where it was written.

\python{INFER}, the default of \python{context} and \python{origin}, means that the value is taken from the timeline the stage is eventually applied to. Times are interpreted relative to prior entries, through the \python{origin} mechanism of \cref{sec:origin}, and the \python{context} of the preceding entry is inherited (\cref{sec:context}). This reflects a general principle: only state changes that differ from the previous ones need to be specified. Worked examples of the reference points available, in both time and value, are collected in \cref{tab:originSpecs}.

\begin{table}
    \centering
    \begin{tabular}{l|p{15em}}
         \python{update(AOM__MOT=1,shutter__MOT=1,context="MOT")} & \small Turns on both the MOT AOM and shutter\\
         \hline
          \python{update(AOM__MOT=1,shutter__MOT=1,time=10,context="MOT")} & \small As above, but $10$\,s later\\
         \hline
         \python{update(AOM__MOT=[0.1,1],shutter__MOT=1,context="MOT")} & \small As the first entry, but with the AOM turned on $0.1$\,s later \\
         \hline
         \python{update(AOM__MOT=[0.1,1,"MOT"],AOM__imaging=[0.,1,"AI"])} & \small Two variables with different times and contexts \\\hline
         \python{update(AOM__MOT=[[0.1,1],[0.2,0]],context="MOT")} & \small The MOT AOM is turned on and then off with a single expression \\\hline
    \end{tabular}
    \caption{The input formats for \python{update}; see the API documentation \cite{wignertime_docs} for details. Times are measured from the reference point that the \python{origin} resolves to (\cref{sec:origin}).}
    \label{tab:inputSpecs}
\end{table}

\subsubsection{Ramp}
\begin{figure}
  \centering
  \begin{minipage}[t]{0.32\textwidth}
    \vspace{0pt}
    
      \begin{minted}[fontsize=\scriptsize, breaklines, breakanywhere, frame=none]{python}
def ramp(
    *,
    duration=None,
    time=None,
    time2=None,
    context=INFER,
    origin=INFER,
    origin2=[VARIABLE, 0.0],
    function=
      ramp_function.tanh,
    **vtvc_dict
)
      \end{minted}
    
  \end{minipage}%
  \hfill
  \begin{minipage}[t]{0.68\textwidth}
    \vspace{0pt}
    \centering
    \includegraphics[width=\linewidth]{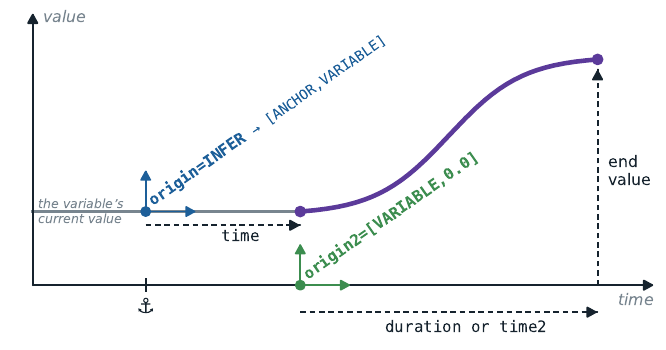}
  \end{minipage}
  \caption{\label{fig:ramp} Programmatic and visual overview of a \python{ramp}. A ramp is defined by two time-value pairs and an interpolating function between them, and therefore requires two origins. \python{origin} places the starting time, as for \python{update}, and the start value is always that of the variable at that instant; \python{origin2} places the ending pair, its default referring the end \emph{time} to the start of the same variable. The end \emph{value} is absolute – the ramp finishes at the value given. The \python{duration} argument is a convenience alias: since the second point’s temporal origin is almost always the first point, \python{duration} and \python{time2} are equivalent in the common case. Where both are specified, \python{time2} takes precedence.}
\end{figure}

While \python{update} places variables at discrete time points, analog variables often require smooth transitions to avoid discontinuities. The \python{ramp} function defines such a transition between two time-value pairs – a start and an end point – via a user-specified function, defaulting to hyperbolic tangent (\cref{fig:ramp}).
 
\python{ramp} extends \python{update} by a transition function, and by an end point for each variable in place of a single point. Its start value is not written: a ramp starts where its variable is, since a start that differed from the current value would be a jump hidden inside the ramp. In practice, ramps are almost always defined by an end value and a duration, and the argument list is designed to make this as concise as possible.

By default, \python{ramp} stores only the start and end time-value pairs along with a function reference – it does not immediately evaluate the function over the domain. This keeps the signal-to-noise ratio high and the memory footprint low in the physics layer. The timeline is only fully expanded when \python{expand} is called, which is normally automated before hardware conversion. The function is then sampled at the cycle period of the controller, unless it fixes a resolution of its own: a ramp given \python{function=functools.partial(ramp_function.tanh, time_resolution=1e-4)} is sampled every 0.1\,ms, which keeps a long ramp from occupying every cycle it spans.
 
Like \python{update}, \python{ramp} inherits the \python{context} of the preceding entry by default (\cref{sec:context}), but this can always be overridden explicitly.
 
\paragraph{What a ramp refuses} A ramp must end after it begins: a zero or negative \python{duration} raises an error. A ramp whose value does not change is not an error but a hold, and occupies time like any other ramp. Since a ramp starts from the current value of its variable, ramping a variable that has not yet been set also raises an error, and so does starting a ramp of a variable while another ramp of it is in progress, since the value it passes through at that instant is in no row of the timeline. One ramp may start exactly where another ends.
 
Examples are given in \cref{tab:rampExamples}.
 
\begin{table*}
\begin{tabular}{p{20em} | p{15em}}
\begin{pythonTable}
ramp(
    coil__MOT_lower__A=0.0,
    coil__MOT_upper__A=0.0,
    duration=duration,
    context="MOT")
\end{pythonTable}
& \small The typical usage: both MOT coils are ramped down over \python{duration}, e.g. for a polarization-gradient cooling stage. End values are specified per variable; all other options are shared. The starting time is inferred from the previous timeline via \python{origin}.
\\\hline
\begin{pythonTable}
ramp(lockbox__MOT__MHz=[500e-3,0.0])
\end{pythonTable}
& \small For simpler cases, a variable’s duration and end value can be supplied as a list.
\\\hline
\begin{pythonTable}
ramp(lockbox__MOT__MHz=[500e-3, 0.0,
    "finalization"])
\end{pythonTable}
& As above, but opening a new \python{context} (see \cref{sec:context}).
\\\hline
\begin{pythonTable}
stack(
    update(lockbox__MOT__MHz=5.0),
    ramp(lockbox__MOT__MHz=0.0,
         duration=0.05))
\end{pythonTable}
& \small A ramp from a value other than the current one: the jump is an \python{update} before the ramp, where it shows in the timeline.
\end{tabular}
\caption{\label{tab:rampExamples} Examples of \python{ramp} usage. In each case the starting time is left to the \python{origin} mechanism, the ramp starts from the value its variable has at that instant, and the end value is the value the variable is ramped to.}
\end{table*}

\subsubsection{Anchor}
\label{sec:anchor}
An \emph{anchor} is a special kind of \python{origin} – a virtual reference time that can be placed in the timeline using:
\begin{minted}[fontsize=\footnotesize, breaklines, frame=none, linenos=false]{python}
def anchor( time, *, context=INFER, origin=INFER )
\end{minted}
Anchors are useful because key experimental time instants are often necessarily virtual. An anchor marks a moment that matters physically but that no device update records. The commonest case is a stage that ends in a waiting period: a MOT collection or a molasses stage finishes not because something is switched, but because enough time has passed. Nothing is commanded at that instant, so without an anchor there is no row to refer to, and the following stage would have to be placed relative to whichever update happened to come last – an accident of how the stage was written rather than a statement about the experiment. The anchor makes the waiting period explicit and gives the instant a name that later stages can chain from via the \python{origin} mechanism.
 
An anchor is simply a special row in the timeline – one whose \python{variable} entry does \emph{not} correspond to a connection and contains the designated anchor symbol (\faAnchor~ by default). It can therefore be added using the convenience function above or by any other means of adding a row. It is recommended that all user-defined stages end with an anchor whose \python{context} denotes the stage.

Note that \python{time} is required, and is a displacement from whatever the \python{origin} resolves to rather than an absolute time. Since the default origin is the most recent anchor where one exists, \python{time} is normally a duration measured from the end of the \emph{preceding stage} – which is precisely what allows a stage to write rows past its own closing anchor, as \python{optical_pumping} does when it reinitializes its shutters, without dragging the following stage along with them. Where the intention is instead to mark the end of everything written so far, that is a different instant and is asked for explicitly:
\begin{minted}[fontsize=\footnotesize, breaklines, frame=none, linenos=false]{python}
anchor(0.1, origin=LAST)  # 0.1 s after the last entry in the timeline
anchor(0.1)               # 0.1 s after the most recent anchor
\end{minted}
The two coincide only when the last stage ended with an anchor and wrote nothing after it. In the example of \cref{sec:demonstration}, \python{optical_pumping} reinitializes its shutters 0.1\,s after its anchor, so throughout magnetic trapping the last entry lies about 0.1\,s after the most recent anchor. The anchor of the final stage, a second later, makes the two coincide again. Which of the two is meant should therefore always be stated.
 
\subsubsection{From a stage to a timeline}
\label{sec:to_timeline}
\begin{minted}[fontsize=\footnotesize, breaklines, frame=none, linenos=false]{python}
def to_timeline( stage, onto=None )
\end{minted}
A stage is applied to the timeline given as \python{onto}, or to an empty one, and the result is a new timeline. No function modifies the timeline it is given. In contrast to most other scientific software, every function returns new data instead, which is the functional paradigm in action, and all user-defined functions are recommended to follow the same pattern. The principal advantage is the natural ability to construct pipelines, a design principle that is particularly exploited by the \python{stack} function (\cref{sec:stacking}).

At the start of an experiment there is nothing to refer to or to inherit from, so the first rows are placed in absolute time and must name their context: every row of a timeline has one.

The \python{onto} argument places a stage into an existing timeline, even a finished one (\cref{sec:interweaving}). Since \python{to_timeline(b, onto=to_timeline(a))} is the same timeline as \python{to_timeline(stack(a, b))}, a base timeline can be built once and each variation of a parameter scan placed onto it (\cref{sec:parameter_scan}).

\subsection{Context}
\label{sec:context}
The fourth core column, \python{context}, labels the experimental stage to which a row belongs, e.g. \python{"MOT"} or \python{"optical_pumping"}. It carries no timing information of its own and is never sent to the hardware, but it does three distinct kinds of work.
\begin{enumerate}
\item It is documentation that survives into the data. An assembled timeline records not only what happened and when, but which part of the experiment each update belonged to – which is what makes an archived run readable a year later, by someone who did not write it. The display conveniences use it to group and annotate a plot (\cref{fig:timeline__example}), and \python{timeline.context_info} reports, for each context, the set of variables it touches and its temporal extent, which is often the quickest way to check that a composed timeline is what was intended.

\item A context is addressable as an \python{origin} (\cref{sec:origin}). A stage is thereby a \emph{named} region of the timeline, and other operations can be placed relative to it without any absolute instant being computed – the mechanism underlying interweaving (\cref{sec:interweaving}).

\item A backend may reserve particular context names for its own purposes. In the ADwin case, \python{"ADwin_LowInit"} and \python{"ADwin_Finish"} mark rows that are actuated outside the timed sequence – on initialization, and on termination including interruption – so that the apparatus is left in a defined state however the run ends (\cref{sec:demonstration}). Such rows belong to no instant of the run, and the timeline says so: the initial state is written at \python{time=-math.inf} and the final state at \python{time=math.inf}. Nothing is placed relative to either, and a stage written after the initial state starts at absolute zero.
\end{enumerate}

Contexts are inherited rather than repeated. Where \python{context} is not given explicitly, \python{update}, \python{ramp} and \python{anchor} adopt the latest context of the timeline they extend, so a stage need be named only once. The initial and final states pass on no context, so a row written after them names its own, or takes that of the last stage of the run. The same principle operates from above: a \python{context} passed to \python{stack} is forwarded to all of its constituents (\cref{sec:stacking}), so a single \python{context="MOT"} labels every row that stack produces, except those of a constituent that names its own. This is the general rule that only what differs from the preceding state need be stated. Every row of a timeline has a context, stated or inherited.
 
\subsection{Combining stages}
The core functions described above are designed to be composed into \emph{pipelines}: chains of stages, each extending the timeline left by the one before. The \python{stack} function and its derivatives make this composition convenient and readable.

\subsubsection{Stacking}
\label{sec:stacking}
A stage is a function of a timeline, with all its other arguments already supplied. This allows experimental stages to be defined independently of any specific timeline – relative to their own start and end, without reference to absolute time. It is only when \python{to_timeline} applies the composition that each stage is applied in turn, each extending the timeline returned by the one before, so that the chain resolves to a single timeline.

This is central to the user experience of Wigner Time. Stages such as optical pumping or absorption imaging can be defined once, generically, and inserted anywhere within a specific timeline. Such building blocks are composed using \python{stack}:
\begin{minted}[fontsize=\footnotesize, breaklines, frame=none, linenos=false]{python}
def stack(*stages: Callable, **kws) -> Callable
\end{minted}
A \python{stack} of stages is itself a stage, so stacks can be nested and can grow indefinitely, until \python{to_timeline} makes a timeline of the whole.

All keyword arguments passed to \python{stack} are forwarded to its stages, reducing duplication. They serve as defaults: a stage that states the same argument keeps its own. This is particularly convenient for assigning a shared \python{context}, e.g.
\begin{minted}[fontsize=\footnotesize, breaklines, breakanywhere, frame=none]{python}
stack(
    update(...),
    ramp(...),
    context="MOT"
    )
\end{minted}
Note that stages are written in execution order, the immediate advantage being legibility over nesting one call inside another. The deeper value, however, is in the modularity it enables.

This is best illustrated by example. Suppose we want to estimate the number of atoms in an atomic cloud. The procedure requires configuring a combination of lasers (MOT and repump), setting electromagnet drive currents, collecting atoms in a magneto-optical trap for 0.1\,s, reconfiguring the lasers, and recording fluorescence on a CCD.

Breaking this down into reusable operations, the MOT stage is clearly central, and can be defined as a reusable function. It opens the appropriate shutters and brings the electromagnets to the stage’s currents, by a smooth ramp to minimize self-induction:
\begin{minted}[fontsize=\scriptsize,frame=none]{python}
def MOT(duration, lower_current, upper_current):
    return stack(
        update(shutter__MOT=1, shutter__repump=1),
        ramp(coil__MOT_lower__A=lower_current, coil__MOT_upper__A=upper_current,
             duration=0.1),
        anchor(duration),
        context="MOT"
    )
\end{minted}
This returns a stage that can be slotted into any later timeline. Its arguments are best named explicitly rather than collected in \python{**kwargs}: \python{update} and \python{ramp} read any keyword they do not recognize as a variable name (\cref{sec:functions}), so a misspelled argument would become a new variable without any error. The exception is the default state of the apparatus, where reading unrecognized keywords as variables is exactly what is wanted (\cref{sec:forwarding}).

Similarly, triggering a camera is a useful abstraction, parameterized by the time instant and the duration of the exposition:
\begin{minted}[frame=none]{python}
def trigger_camera(time, exposure, context, origin):
    return update(trigger__camera=[[time,1],[time+exposure,0]],
                  context=context, origin=origin)
\end{minted}
This pattern – a thin wrapper around \python{update} – covers the vast majority of digital operations. Unlike \python{MOT}, the camera trigger is \emph{placed}: its \python{origin} has no default, so the point it is measured from is always named where it is used (\cref{sec:interweaving}).

The initial conditions are defined separately:
\begin{minted}[fontsize=\scriptsize, frame=none]{python}
def init():
    return stack(
        update(
            AOM__MOT=1, AOM__repump=1, shutter__MOT=0, shutter__repump=0,
            trigger__camera=0,
            coil__MOT_lower__A=1, coil__MOT_upper__A=1,  # kept energized between runs
        ),
        anchor(0.0),
        context="initialization"
    )
\end{minted}
As the first rows of the experiment, these are placed at absolute zero. It is advisable to always define an explicit starting condition for every connection in this way.

The complete timeline is then assembled through composition:
\begin{minted}[frame=none]{python}
timeline = to_timeline(stack(
    init(),
    MOT(0.1,1,1),
    update(AOM__repump=0),
    trigger_camera(0.0,1e-3,"imaging",origin="MOT")
))
\end{minted}
The additional \python{update} switching off one laser before imaging slots naturally between custom functions without friction.

The brevity of this declaration follows from a series of design decisions. Since the stages keep the default origin, operations are chained one after the other, and the camera trigger, placed at the end of the MOT stage, falls in line with them. Since anchors are used, every \python{time} argument becomes $\Delta t$ from the previous anchor. The result is code with a high signal-to-noise ratio that nonetheless compiles to a concrete table of time-value rows in absolute terms.

A further useful function for stacking is \python{cascade}:
\begin{minted}[fontsize=\footnotesize, breaklines, frame=none, linenos=false]{python}
def cascade(*fs: Callable, **kws) -> Callable
\end{minted}
This serves a similar purpose to \python{stack} at the outer layer of composition, but with the added feature that keywords are selectively forwarded to the underlying functions by name prefix.

The two take their stages differently, although their syntax does not show it. \python{stack} takes stages that have \emph{already been called} – \python{stack(init(), MOT(duration=15))} – where every argument has been supplied and what remains is a function of the timeline alone. \python{cascade} takes the stage functions \emph{themselves} – \python{cascade(init, MOT, MOT_duration=15)} – and calls them, having routed each keyword to its stage. A stage function therefore reaches \python{cascade} bare and reaches \python{stack} called, and the two are not interchangeable. Like \python{stack}, \python{cascade} returns a stage, so a \python{cascade} is itself composable. Once a good set of modular functions with default arguments is established, it is often useful to adjust specific nested parameters from a single point of contact. This is achieved by prefixing keywords with the function name, e.g.
\begin{minted}[frame=none]{python}
cascade(
    init,
    MOT,
#
    MOT_duration=5.0,
    MOT_lower_current=-1.0,
    MOT_upper_current=-0.98
)
\end{minted}
Routing is strict: a keyword that matches no stage, or none of the parameters of the stage it matches, raises an error – unless that stage accepts arbitrary keywords, as \python{init} and \python{finish} do (\cref{sec:forwarding}). Where one stage name begins another, as \python{MOT} begins \python{MOT_detuned_growth} in \cref{sec:demonstration}, the longer name is tried first.

\subsubsection{Interweaving}
\label{sec:interweaving}
Stacking is a powerful technique for composing linear timelines, where each stage follows the previous one. In practice, non-linear timelines are also useful – where an operation is anchored not to the end of the preceding stage but to some earlier reference point.
 
Returning to the example above (\cref{sec:stacking}), the \python{trigger_camera} function need not follow linearly from the MOT stage. Given \python{origin="initialization"} instead, the camera trigger is placed relative to the initial anchor. Such \emph{interweaving} is particularly valuable for inserting diagnostic operations into an existing timeline without restructuring it.

The timeline need not even be unfinished. Taking the complete \python{timeline_demo} of \cref{sec:demonstration}, with the camera trigger added to the apparatus as in \cref{sec:forwarding}, a single call triggers the camera 2\,ms after the end of the molasses stage:
\begin{minted}[frame=none]{python}
timeline = to_timeline(
    trigger_camera(2e-3, 1e-3, "imaging", origin="molasses"),
    onto=timeline_demo)
\end{minted}
The exposure falls during magnetic trapping, well before the final ramps, and every row already in \python{timeline_demo} stays where it was.

This gives a practical criterion for what a stage should declare. A stage that is meant to be \emph{placed} takes \python{origin}; a stage that forms a link in a causal chain does not. The preparation sequence of \cref{sec:demonstration} is of the second kind – each of \python{MOT}, \python{molasses}, \python{optical_pumping} and \python{magnetic_trapping} acts on the physical state the previous one left behind, so there is no meaningful freedom in where it goes, and an \python{origin} argument would exist only to be given wrongly. The diagnostics of \cref{sec:forwarding} are of the first kind, and every one of them declares it. The signature therefore says which a stage is, and that is worth preserving as the operation layer grows.
 
In \cref{sec:parameter_scan}, the same mechanism places an imaging stage at a series of delays for a time-of-flight measurement.
 
\section{ADwin back-end}
\label{sec:adwin}
While the vast majority of Wigner Time is concerned with the design, creation and manipulation of timelines, at some point it is necessary to translate these to a real-time hardware system.
 
The ADwin real-time data acquisition and control device is one such system – a self-standing unit combining hardware, an IDE, a language, and a compiler \cite{ADwin}. Having gained popularity in AMO physics within the last decade, ADwin has been used as a back-end in the NQontrol platform \cite{darsow2020nqontrol}, which provides multi-channel digital feedback loops for quantum-optical experiments. Although it currently competes with other options, such as the Sinara/ARTIQ control system \cite{kasprowicz2020artiq,bourdeauducq2016artiq} and NI’s compactRIO systems, ADwin is the only system that has been tested by the authors and so is the only one for which integration software is provided. As outlined below, however, so little code is needed for the integration that users should not have difficulty building bridges to other manufacturers.
 
Two properties are worth emphasizing, precisely because they belong to the backend rather than to Wigner Time. First, the determinism of the output – timing jitter and long-term drift – is set by the timing hardware and its clock-distribution chain, not by the layer that generates the timeline; ADwin systems can be referenced to an external clock, so a laboratory-grade frequency standard can be used wherever the experiment demands it. Second, ADwin is modular, so which channel types are available is a question of which modules are installed, not of the control software. Wigner Time inherits both properties here, and would inherit the corresponding properties of any other backend – which is what back-end agnosticism means in practice.
 
Wigner Time comes with a subpackage, \python{wignertime.adwin}, that provides many conveniences. Some of them know only what is particular to ADwin and touch no hardware: the display module, for instance, which knows the contexts that ADwin reserves and draws the initial and final states either side of the run. Others, like those in \python{wignertime.adwin.core}, interact directly with the hardware and rely on the official ADwin Python support \cite{adwin_pypi}.
 
In practice, few of these functions are necessary on a day-to-day basis. As demonstrated in \cref{sec:demonstration}, the main function required is \python{wignertime.adwin.core.upload}, which converts a timeline at the cycle period read from the controller itself and writes the result into the controller’s memory. It returns a record of what was written, which can be archived alongside the data.
 
The abstraction away from ADbasic is one of the central motivations for Wigner Time. For open-loop control, the official ADwin language is unnecessarily unwieldy: although built for real-time operation, its lack of modularity makes it difficult to maintain in the constantly evolving environment of a typical quantum optics research group. Lifting the experimental logic out of ADbasic is not only a practical convenience – it also reduces cycle time and hence increases the achievable temporal resolution, since any extra logic in the real-time program requires extra computation. To this end, the essential ADbasic code for running daily experiments, ignoring constants and variable definitions, can be reduced to:
\begin{minted}[fontsize=\footnotesize,frame=none]{basic}
sub processUpdates(cc)
  ' analog
  if ( (analogIdx <= analogArrayDim) and (data_10[analogIdx] = cc) ) then
    do  
      p2_dac(data_11[analogIdx],data_12[analogIdx],data_13[analogIdx])
      inc analogIdx
    until ( (analogIdx > analogArrayDim) or (data_10[analogIdx] > cc) )
  endif
  
  ' digital
  if ( (digitalIdx <= digitalArrayDim) and (data_20[digitalIdx] = cc) ) then
    do
      p2_digout(data_21[digitalIdx],data_22[digitalIdx],data_23[digitalIdx])
      inc digitalIdx
    until ( (digitalIdx > digitalArrayDim) or (data_20[digitalIdx] > cc) )
  endif
endsub.
\end{minted}
 
As can be seen, the role of ADwin has been reduced to a single concept – “at this
cycle, write the values listed for it” – written once each for analog and digital
connections. It does not compare values: which ones have changed has already been
decided before the upload, as described below. The
loop therefore costs two comparisons per channel group, with early exit, and no
arithmetic whatsoever. This is what permits the 1\,µs cycle time quoted in
\cref{sec:intro}: the previous implementation evaluated, on every cycle and for
every ramp in progress, an expression of the form
\python{Tanh(...)/TwoTimesTanhOfTheTanhInterval} and was limited to 5\,µs as a result.

The arrays uploaded to the controller are prepared in the same spirit. Ramps are expanded at the cycle period of the controller, unless a ramp fixes a coarser resolution of its own (\cref{sec:functions}), and converted to DAC codes, and every row that would leave a channel at the code it already holds is then removed. What remains is one row for each change of code, at the first cycle at which that change occurs. Kowalski \textit{et al.} \cite{kowalski2023entangleware} treat ramps in the same spirit, timing analog transitions at the instants when the DAC code changes rather than sampling the trajectory at a fixed rate. A ramp is therefore as smooth as the hardware allows, the controller receives no more rows than are needed to describe it, and no time resolution has to be chosen.

The controller also acquires data. A variant of the sequencer plays the same arrays and, in addition, records one analog input over a window of the run, given in seconds of the timeline. \python{wignertime.adwin.adc.run} takes the record returned by \python{upload} together with the window, and returns the samples with their times on the timeline’s own axis, so that a recording can be read against the timeline that drove it.

\medskip
How the controller divides its outputs between a run and a manual console is described in \cref{sec:adwin_operation}. For further details, consult the API documentation \cite{wignertime_docs}.

\section{Discussion}
\label{sec:discussion}
By eschewing both the object-oriented paradigm and low-level proceduralism, Wigner Time presents a data-oriented design that is more concise and less coupled to implementation than alternative software, while remaining precise and extensible. This section reflects on the most distinctive aspects of this design and considers the future outlook.
 
A first example of the resulting user experience is the exceptionally high signal-to-noise ratio in code legibility. As seen in \cref{sec:demonstration}, realistic code is almost exclusively focused on intent rather than implementation – it answers the \emph{what?} rather than the \emph{how?} This is largely achieved by a combination of hierarchical defaults, heavy use of variable arguments, and function chaining – in short, because it follows a functional paradigm.
 
Such chaining, as exploited by the \python{stack} function, is itself made possible by the general avoidance of statefulness: the output of each function is entirely dependent on its arguments and not on any side effects or global variables. This brings the added benefit that the software is much easier to test than its peers, and so can be made flexible without losing robustness.
 
Flexibility is perhaps the stand-out feature of Wigner Time. By catering heavily for variable arguments, both fixed-order and keyword-based, the vast majority of code can be reduced to understanding and using only four functions: \python{update}, \python{ramp}, \python{stack} and \python{to_timeline}. The first three describe the experiment, and the last makes a timeline of the description. A single call can set any number of variables, and parameters kept in a dictionary can be passed whole by unpacking, e.g. \python{ramp(duration=1e-3, **my_dictionary)} – a further advantage of the Python keyword system. The limit lies in the hardware: every row due at a given cycle must be written within that cycle, and the arrays on the controller have a fixed capacity. \python{upload} refuses a timeline that exceeds the capacity before writing anything, and a run in which the controller fell behind its cycles is reported as an error once it has ended.
 
Flexibility was also the driving force behind the \python{origin} functionality (\cref{sec:origin}). It became apparent that although absolute time is the desired end goal, relative time is the desired method. Wigner Time therefore combines both, allowing linear stacking with the default \python{origin}, interweaving when \python{origin} is a context, and even negative time around a reference point when using non-trivial origins. Furthermore, by deferring ramp expansion, time resolution only comes into play when passing the timeline to the hardware, or optionally when plotting. Because timeline construction is a pure function of its parameters, the same composition extends without modification from a single shot to a parameter scan, and from there to closed-loop operation: an optimizer that consumes the analysis of one shot and proposes the parameters of the next requires no new machinery, only a different source of parameters, where labscript provides a dedicated component for the purpose \cite{starkey2013labscript}. A worked example is given in \cref{sec:parameter_scan}.
 
It is worth noting that parts of this design have been arrived at independently elsewhere. The Entangleware Sequencer \cite{kowalski2023entangleware}, developed concurrently and for entirely different hardware, converges with Wigner Time on three points: that the instruction stream should be precompiled rather than interpreted in-loop, so that almost no logic remains in the real-time system; that a Python front end should organize an experiment hierarchically, with lower-level routines close to the hardware and mid-level ones named after experimental stages; and that the coexistence of relative and absolute timing is a first-class design problem rather than an implementation detail. Their \python{abs} and \python{rel} methods are, to our knowledge, the closest published analogue to our \python{origin} mechanism. Such convergence is encouraging: it suggests that these are properties of the problem rather than of our taste.
 
The treatment of time nevertheless differs instructively, and the comparison is naturally three-way. In Cicero \cite{keshet2013distributed}, sequence steps are strictly relative, each beginning when the previous one ends – which, as Kowalski \textit{et al.} observe, makes interwoven operations with differing start times and durations awkward to express. Entangleware resolves this by admitting absolute times alongside relative ones, at the cost that a genuinely interwoven operation must be placed by arithmetic performed by the user: in their published example, a digital pulse is scheduled at a hand-summed offset from the start of the sequence. Wigner Time takes a third route, in which the reference point is \emph{named} rather than computed. An \python{anchor} (\cref{sec:anchor}) records the instant that matters physically, and the arithmetic that would otherwise be written out by hand is resolved against that name when the timeline is assembled. Sub-sequences therefore remain relative to their own internal logic, and independent of where they are eventually inserted.
 
The deeper divergence, however, is not about time at all, but about whether the experimental description exists as an inspectable object. In class-based designs of this kind, constructing the sequence \emph{is} the act of enqueueing it: the user’s methods call output primitives directly, timing state is advanced by mutation, and the instruction stream is accumulated between explicit “build” and “run” calls. The description is thus consumed as it is produced, and there is no intermediate representation to plot, filter, difference against yesterday’s run, serialize to disk, or hand to a collaborator who does not share the hardware. In Wigner Time the timeline is a value, and every one of those operations comes for free – \cref{fig:timeline__example} is generated from the same object that is subsequently uploaded to the hardware.
 
A related consequence concerns where hardware detail resides. Where the description is a program, module and channel numbers – and the conversion from physical units to DAC codes – tend to appear inside the code that describes the physics, so that rewiring a connection or recalibrating a device means editing experimental logic. The \python{device} and \python{connection} tables (\cref{sec:hardwarelayer}) exist precisely to prevent this: the physics is written in named variables and physical units, while calibration and wiring are each declared once, elsewhere, and changed independently.
 
A note on scope is also in order. Wigner Time addresses the composition of hardware-timed sequences, and deliberately does not attempt the instrument-orchestration and measurement-logic role filled by suites such as Qudi \cite{binder2017qudi}, which separates hardware abstraction, experiment logic and user interface into distinct layers and ships ready-made measurement protocols; the layering philosophies are nonetheless comparable.
 
The final concern, downstream of flexibility, is how the library can evolve as demands change. One present limitation is the lack of integration with hardware outside of ADwin; although it would not take a developer long to bridge the gap, the absence of ready-made conversions for ARTIQ or NI devices is a hiatus nonetheless. A second gap of the same kind concerns peripherals that are not driven by a voltage but programmed over a serial interface – direct digital synthesizers being the canonical example, where setting a frequency means clocking out a tuning word across several digital lines, followed by an update strobe. Nothing about this is foreign to the physics layer, since a frequency is simply a variable with a unit like any other; what is required is an expansion in the conversion step that turns one such row into the corresponding bit-level rows. This is structurally the same operation that \python{expand} already performs for ramps, and we expect it to be added in that form rather than as a special case.
 
The positive future outlook is largely based on the careful attention to layered abstraction. The base layer can hardly get simpler, and users who do not require more flexibility can simply stay at the level of table manipulation. At the same time, it is straightforward to add higher layers of abstraction as needed: new stages or experimental chains can be wrapped in a function and, as seen in \cref{sec:stacking}, interleaved with the core functions. As an example, the camera trigger function from that section could easily be expanded by a few lines of code to cater for the pulsing of all digital switches an absorption imaging requires.
 
A less visible but equally consequential benefit concerns the longer term, over which laboratory software is written by successive students and rarely by its original author. Where the description is a program over a global clock, tunable quantities tend to accumulate in numbered global slots, mapped to names in a header file and set externally at run time; the parameters that produced a given dataset are then not in the source at all, and which of them belong to which stage cannot be recovered by reading. Specifications needed at more than one point in the experiment fare no better. The apparatus’ default state, for instance, is required both before a run and after it, and in our previous implementation it was written twice, in the initialization and finalization sections, from which the two copies had quietly diverged – disagreeing on several shutters and omitting others altogether. In Wigner Time the same specification is a function, \python{default_state}, applied at both ends of the experiment and differing only in an argument, \python{MOT_ON}, that records the one difference that is actually intended. This is what we mean by curation: not that the software is easier to write, but that its intent survives being read by someone else a year later.
 
The data-oriented description makes version control worth applying, because the diff is meaningful.
 
\section{Conclusion}
\label{sec:conclusion}
We have introduced a new design and Python software implementation for real-time,
open-loop control systems, in which the experimental procedure is decoupled from
the timing hardware by being represented as data rather than as a program.
 
Scientific software in this domain is remarkably unstandardized: each lab typically develops its own tooling on an ad hoc basis, usually without version control, and with little institutional reward for publishing the result. The systems surveyed in \cref{sec:intro} are the visible exceptions rather than the rule, and even among these, some of the most widely reused – the control system of Meyrath and Schreck prominent among them \cite{meyrath_schreck_control} – exist only as project pages rather than in citable form. Laboratory programs consequently lag far behind modern computing trends – a situation that also makes it difficult to cite prior work.
 
It is in this context that Wigner Time offers tools that are unusually composable and portable, enabling high-performance hardware control while making a modern functional paradigm available to physics labs internationally.
 
\section*{Acknowledgements}
T. W. Clark would like to acknowledge and thank the informal contribution of the Clojure community in general, as well as the Clojure data science community in particular, who have been working to bring data-oriented programming into the mainstream.

A systematic review of the codebase was carried out with the assistance of Anthropic’s Claude Code. The same tool has subsequently been used for rectification, verification, and documentation. The design decisions, and the final judgement on each finding, remained with the authors.

\paragraph{Code availability}
Wigner Time is developed openly at \url{https://github.com/wignerquantumoptics/Wigner_Time}, under the GNU General Public License, version 3 or later, and its documentation is at \url{https://wignerquantumoptics.github.io/Wigner_Time/}. It is written in Python and runs on any operating system; the real-time program on the controller is written in ADbasic. It requires Python 3.10 or higher, pandas, SciPy, funcy and munch, and optionally the ADwin Python package \cite{adwin_pypi}, for running experiments on ADwin hardware, and matplotlib, for display.

\paragraph{Author contributions}
TWC, BS and AV designed Wigner Time, and TWC and AV implemented it. AD, ÁK, DN, AS, DV and PD made it work in a laboratory. They ran the two cold-atom experiments with it daily for more than two years, and each version was tested on their apparatus, which is the only place where the hardware back end can be verified. What they needed to express in the course of real experiments – stages placed relative to a named point of a running sequence, diagnostics added to an existing apparatus, scans over a parameter – is what the operation layer of this article answers, and their reports of failures and their design ideas shaped the package throughout. A complete experiment run by DV on one of the setups is part of the package’s test suite. AV and TWC drafted the paper, and all authors read and approved the final manuscript.

\paragraph{Funding information}
This research was supported by the Hungarian National Research, Development and Innovation Office (Grant Nos. 2022-2.1.1-NL-2022-00004 and 2025-3.1.1-ED-2025-00011), the ERANET COFUND QuantERA programme (MOCA 2019-2.1.7-ERA\_NET-2022-00041), the QuantERA II Programme (V-mag 2024-1.2.2-ERA\_NET-2024-00012), and by the Swiss National Science Foundation (Grant No. 230870). AD and TWC acknowledge support from the János Bolyai research scholarship of the Hungarian Academy of Sciences.

\begin{appendix}
\numberwithin{equation}{section}
 
\section{The origin mechanism in full}
\label{sec:origin_full}
\Cref{sec:origin} describes the two cases that account for almost all use of the \python{origin} keyword. This appendix gives the complete specification, which exists to cover the remainder: interweaving against explicit variables, diagnostics, absolute placement, and programmatic construction.
 
The \python{origin} argument accepts
\begin{itemize}
    \item a \python{float}: all times are offset by this number
    \item a \python{str}: the name of a variable or of a context, resolved against the timeline (see below)
    \item one of the tags \python{ANCHOR}, \python{LAST} and \python{VARIABLE} (see below)
    \item a pair, following the same options for each: sets the numerical zero of the time and value columns independently
\end{itemize}
The default, \python{INFER}, does not suppress the feature, but selects the most recent anchor if one exists, and the most recent entry otherwise. Should neither be available, as at the start of an experiment, the new entries are placed in absolute time. The two slots are completed independently, so a partially stated origin keeps the default for the slot it leaves out – \python{origin="molasses"} passed to \python{ramp} takes its time from that context and its value from the variable itself. \python{None} means the same as \python{INFER}, as the whole origin or in one slot, and absolute placement is requested with \python{0.0}.

Three tags have dedicated meanings:

\python{ANCHOR} – time is relative to the most recent anchor entry (\cref{sec:anchor}), i.e. a variable whose name contains the \faAnchor~symbol, or the configured equivalent in \python{wignertime.config}.

\python{LAST} – time is relative to the entry with the highest time recorded so far in the timeline.

\python{VARIABLE} – each variable is placed relative to its own most recent entry, independently of the others. This is the one tag admissible in either slot.

\begin{figure}
  \centering
  \includegraphics[width=0.95\linewidth]{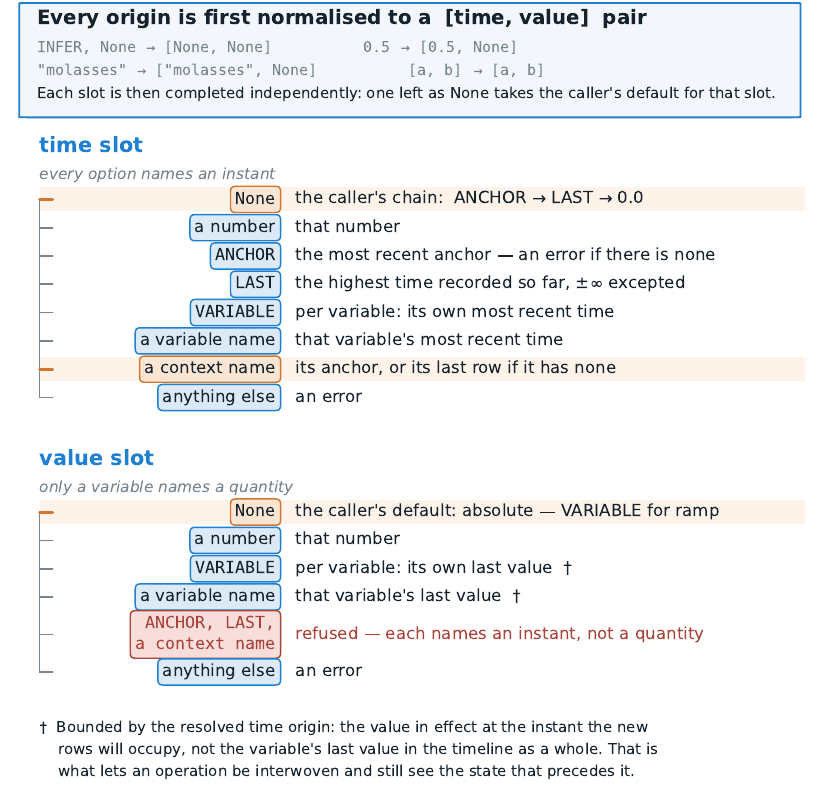}
  \caption{\label{fig:origin} Resolution of an \python{origin}. The two slots admit different vocabularies because they ask different questions: the time slot asks \emph{when}, and every option shown names an instant, whereas the value slot asks \emph{how much, of what}, and only a variable names a quantity. \python{ANCHOR}, \python{LAST} and a context name are therefore refused there – each would resolve to whichever variable happened to hold the row at that instant, and so would answer in the wrong physical units. Nothing is lost, since the intended reading is already expressible as a pair: the value \python{coil__MOT__A} held at the end of the molasses stage is \python{origin=["molasses", VARIABLE]}, the context bounding the lookup in time and the variable naming what is looked up. The two highlighted branches are the cases that occur in nearly all user code: the default, which resolves to the most recent anchor, and the resolution of a context name to that context’s anchor (\cref{sec:origin}).}
\end{figure}

\python{LAST} and \python{ANCHOR} are defined temporally and are admissible in the time slot only, as is a context name; the reason is given in the caption of \cref{fig:origin}.

A string is always a name. If it matches a variable name, e.g. \python{"shutter__MOT"}, the last occurrence of that variable is used. If it matches a context name, e.g. \python{"MOT"}, the origin resolves to the anchor of that context if one is present, and to its final entry otherwise. Such flexibility carries some risk of obscuring the simplicity of the software, but these cases are implemented because they track user intuition; when in doubt, a numeric \python{(time, value)} coordinate is always unambiguous. \Cref{fig:origin} gives the resolution order in full, and \cref{tab:originSpecs} illustrates the possibilities on a single function.

Only rows at an instant of the run are referred to in the time slot. The initial and final states, at $-\infty$ and $+\infty$ (\cref{sec:context}), are not, so a context holding only one of them names no instant, and referring to it raises an error.

Where a value origin is resolved against a variable, the lookup is bounded in time: the value taken is the one in effect at the origin instant, not the variable’s last value in the timeline as a whole. This is what allows an operation to be interwoven into the middle of an existing timeline and still see the state that physically precedes it, including the initial state.

The \python{ramp} function takes a second origin, \python{origin2}, following the same specification and governing the end point of the transition. Its default, \python{[VARIABLE, 0.0]}, refers the end point’s \emph{time} to the start point of the same variable, and its value slot to nothing at all, since the end value of a ramp is always stated. Together that is what makes \python{duration} and a target value a complete specification of a ramp. The start point is the exception: its default is \python{[[ANCHOR, VARIABLE], [LAST, VARIABLE]]}, whose value slot refers the ramp’s starting value to the variable’s own value at that instant. A ramp always starts where its variable is, so this value slot admits \python{VARIABLE} only. Elsewhere no default is value-relative, and value origins are available but are always requested explicitly.

\begin{table}
    \centering
    \begin{tabular}{p{0.57\linewidth}|p{0.4\linewidth}}

    \begin{pythonTable}
update(AOM__MOT=0,
  shutter__MOT=0, time=1)
    \end{pythonTable}
    & \small Turns off both the MOT AOM and shutter $\Delta t=1$\,s after the \emph{latest anchor} in the timeline. If no anchor is found, then after the latest entry in the timeline.\\\hline

    \begin{pythonTable}
update(AOM__imaging=1,
  time=1, origin="molasses")
    \end{pythonTable}
    & \small Activates the imaging-beam AOM $\Delta t=1$\,s after the \python{"molasses"} anchor, or after the latest entry in the molasses context if no anchor is present.\\\hline

    \begin{pythonTable}
update(lockbox__MOT__V=0.5,
  time=1, origin="AOM__imaging")
    \end{pythonTable}
    & \small Sets the lockbox voltage 1\,s after the latest change of \python{AOM__imaging}.\\\hline

    \begin{pythonTable}
update(AOM__imaging=1,
  lockbox__MOT__V=0.5, time=1,
  origin=VARIABLE)
    \end{pythonTable}
    & \small\python{AOM__imaging} and \python{lockbox__MOT__V} are each placed relative to their own most recent entry.\\\hline

    \begin{pythonTable}
update(AOM__imaging=1,
  time=1, origin=0.)
    \end{pythonTable}
    & Sets \python{AOM__imaging} at $t=1$\,s in absolute time (origin at $t=0$).\\\hline

    \begin{pythonTable}
update(lockbox__MOT__V=0.5,
  time=1, origin=[1.0, 4.0])
    \end{pythonTable}
    & \small As above, but with offsets in both time and value: the lockbox voltage is set to 4.5\,V at $t=2$\,s.\\\hline

    \end{tabular}
    \caption{\label{tab:originSpecs} Different ways of using \python{origin}, illustrated with \python{update}, each stage placed onto an existing timeline with \python{to_timeline(stage, onto=timeline)}. The first entry is the default case of \cref{sec:origin} and covers the majority of user code; the second is the resolution of a context name. The remainder are the exceptional cases – explicit variables, per-variable placement, and absolute or numerically offset origins.}
\end{table}

\section{Demonstration: A realistic cold-atom experiment}
\label{sec:demonstration}
\begin{figure}[htb]
  \centering
  \includegraphics[width=1.1\linewidth]{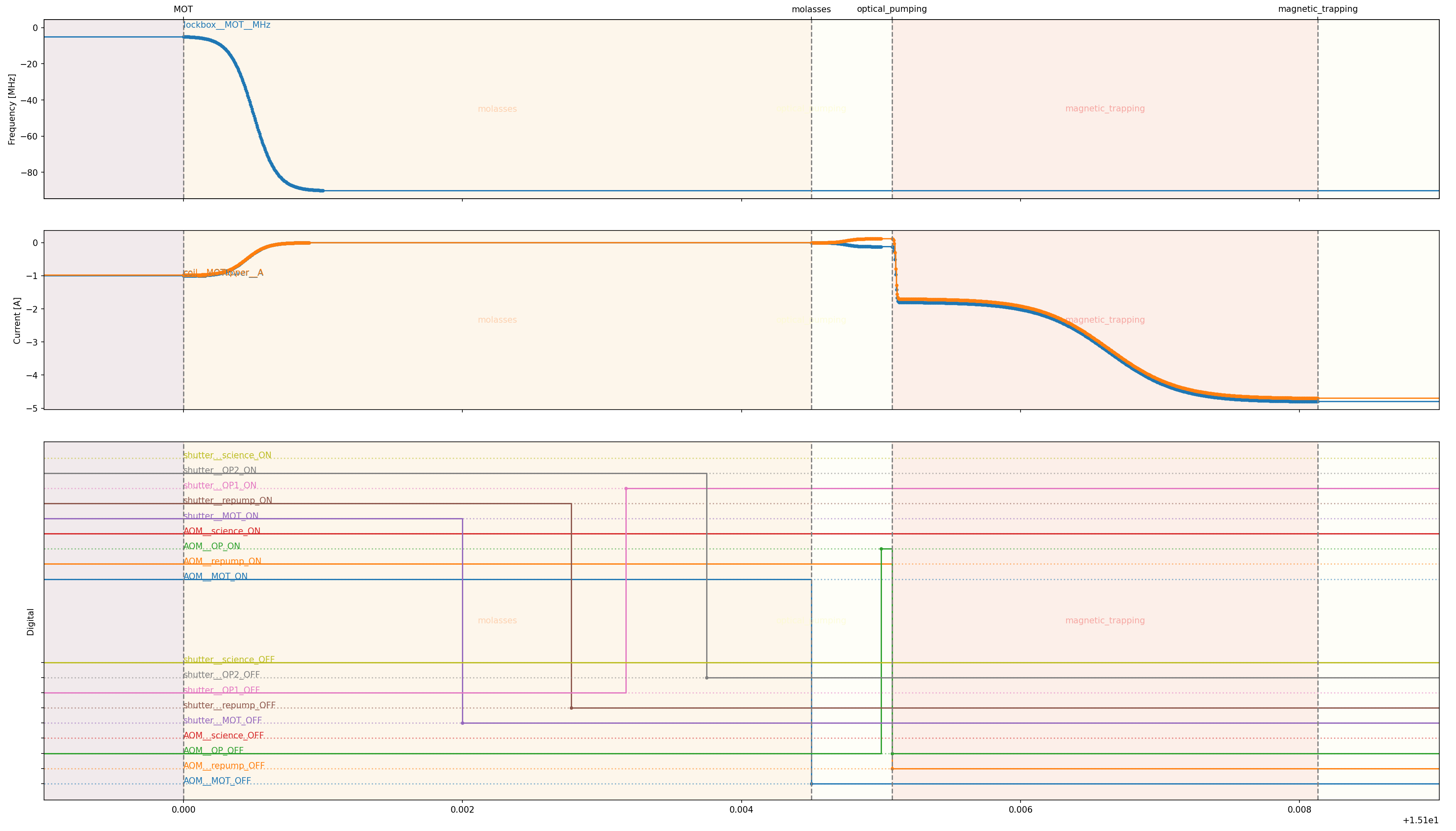}
  \caption{\label{fig:timeline__example} The \python{matplotlib} figure generated by \python{display.quantities(timeline_demo)}, zoomed and cropped for clarity. As can be seen, quantities are automatically separated by unit, with digital variables offset for visual ease. This particular example uses the adwin subpackage, \python{wignertime.adwin}, for system-sensitive defaults.}
\end{figure}

Below, we consider a realistic timing script, including connection definitions, timeline creation and manipulation and subsequent ADwin control. The timeline that it assembles is the one displayed in \cref{fig:timeline__example}.
 
\begin{minted}[fontsize=\mintedsmall, frame=none]{python}
"""
An example implementation of a real experiment, using Wigner Time timelines.
 
As well as providing conveniences, the functions can be used to document the intention and meaning of each stage.
"""
 
import functools
import math

from munch import Munch

from wignertime.adwin import connection as adcon
from wignertime.adwin import core as adwin
from wignertime import file as wtfile
from wignertime import timeline as tl
from wignertime import device
from wignertime import conversion as conv
from wignertime import ramp_function

###########################################################################
#                       Constants and Helpers                             #
###########################################################################

# Connections, devices and constants can be read from a separate file(s) (they won't change much).
# They are all collected together here for demonstration purposes only.

"""
`connections` allows us to label physical links (inputs and outputs) between devices and the timing system.
By using labels that follow a particular regex, configurable as `config.VARIABLE__REGEX`,
we can separate out the design and the implementation of our experiment.
"""
connections = adcon.new(
    ["shutter__MOT", 1, 11],
    ["shutter__repump", 1, 12],
    ["shutter__OP1", 1, 14],
    ["shutter__OP2", 1, 15],
    ["shutter__science", 1, 10],
    ["AOM__MOT", 1, 1],
    ["AOM__repump", 1, 2],
    ["AOM__OP", 1, 31],
    ["AOM__science", 1, 3],
    ["coil__MOT_lower__A", 4, 1],
    ["coil__MOT_upper__A", 4, 3],
    ["lockbox__MOT__MHz", 3, 8],
    ["AOM__science__trans", 4, 8],
    ["dispenser__Rb__A", 3, 3],
)

"""
`devices` stores how to map our physical quantities to an implementation voltage,
as well as specifying the range of values that should be allowed for this variable.
A linear conversion is the factor taking the unit *to* volts, so a device whose permitted range spans
the controller's full output is `10 / limit` – `10 / 5.0` for a coil driven over +/-5 A by a +/-10 V DAC.

These specifications are deliberately separated from `connection`s because they represent
physical properties and conversions that are independent of the particular DAC wiring.
"""
devices = device.new(
    ["coil__MOT_lower__A", 10 / 5.0, -5, 5],
    ["coil__MOT_upper__A", 10 / 5.0, -5, 5],
    ["lockbox__MOT__MHz", 0.05, -200, 200],
    [
        "AOM__science__trans",
        conv.function_from_file(
            "resources/calibration/aom_calibration.dat",
            sep=r"\s+",
        ),
        0.0,
        1.0,
    ],
    ["dispenser__Rb__A", 10 / 5.0, -5, 5],
)


"""
'constants' allow us to store site-specific details that help define our experiment.
"""
constants = Munch(
    lag_MOT_shutter=2.3e-3,
    lag_repump_shutter=2.3e-3,
    OP=Munch(
        lag_shutter_on=1.48e-3,
        lag_shutter_off=1.78e-3,
    ),
)


###########################################################################
#                   Experimental stages                                   #
###########################################################################
# NOTE: The idea behind the function wrapping is that we enclose what will rarely change and expose just those attributes that we are likely to want to vary.


def default_state(MOT_ON=True, **kwargs):
    """
    Starts/leaves the system in a sane state that is appropriate for creating a new timeline

    As a general rule, AOMs are kept on as long as possible to keep them in thermal equilibrium. 
    When needed, we turn them off before the opening of the shutter. The same holds for the coils.
    """
    return tl.update(
        lockbox__MOT__MHz=0.0,
        AOM__MOT=1,
        AOM__repump=1,
        AOM__OP=1,
        AOM__science=1,
        shutter__MOT=MOT_ON,
        shutter__repump=MOT_ON,
        shutter__OP1=0,
        shutter__OP2=1,
        shutter__science=0,
        AOM__science__trans=1.0,
        dispenser__Rb__A=3.8,
        **kwargs,
    )


def init(**kwargs):
    """
    The initial state belongs to no instant of the run: it is placed before it, at minus infinity,
    in `ADwin_LowInit`, a 'special' context that the ADwin system plays before the run starts.
    """
    return default_state(
        time=-math.inf,
        context="ADwin_LowInit",
        **kwargs,
    )


def finish(wait=1, lower_current=-1.0, upper_current=-0.98, MOT_ON=True, **kwargs):
    """
    Safely winds down the system, 'ramping' the analog variables 
    to the “default state” in a given duration by the default `ramp_function`.

    The `anchor` function is used to specify a key time instant, around which other times can be specified.

    `ADwin_Finish` is again a special context.
    It means that the “default state” will be actuated when the run ends,
    even when it is interrupted, and is therefore placed after the run, at plus infinity.
    """
    duration = 1e-2

    return tl.stack(
        tl.anchor(wait),
        tl.ramp(
            lockbox__MOT__MHz=0.0,
            coil__MOT_lower__A=lower_current,
            coil__MOT_upper__A=upper_current,
            duration=duration,
        ),
        default_state(
            time=math.inf,
            context="ADwin_Finish",
            MOT_ON=MOT_ON,
            **kwargs,
        ),
        context="finalize",
    )


def MOT(duration=15, lower_current=-1.0, upper_current=-0.98):
    """
    Creates a magneto-optical trap
    """
    return tl.stack(
        tl.update(
            shutter__MOT=1,
            shutter__repump=1,
            coil__MOT_lower__A=lower_current,
            coil__MOT_upper__A=upper_current,
        ),
        tl.anchor(duration),
        context="MOT",
    )


def MOT_detuned_growth( duration=100e-3, duration_ramp=10e-3, detuning__MHz=-5 ) :
    """
    Final stage of MOT collection with farther detuned MOT beams
    """
    return tl.stack(
        tl.ramp(
            lockbox__MOT__MHz=detuning__MHz,
            duration=duration_ramp,
        ),
        tl.anchor(duration),
        context="MOT",
    )


def molasses(
    duration=5e-3,
    duration_coil_ramp=9e-4,
    duration_lockbox_ramp=1e-3,
    to__MHz=-90,
    delay=0,  # arbitrary delay to shutter for ad hoc compensation of small drifts
):
    """
    Optical molasses cooling with zero magnetic field
    """

    return tl.stack(
        tl.ramp(
            coil__MOT_lower__A=0,
            coil__MOT_upper__A=0,
            duration=duration_coil_ramp,
        ),
        tl.ramp(
            lockbox__MOT__MHz=to__MHz,
            duration=duration_lockbox_ramp,
        ),
        tl.update(
            shutter__MOT=[duration-constants.lag_MOT_shutter + delay, 0],
            AOM__MOT=[duration, 0],
        ),
        tl.anchor(duration),
        context="molasses",
    )


def optical_pumping(
    duration_exposition=80e-6,
    duration_coil_ramp=50e-6,
    i=-0.12,
    delay1=0,
    delay2=0,
    delay_repump=0,  # arbitrary delays to shutters for ad hoc compensation of small drifts
    delay_shutter_reinitialization=0.1,
):
    """
    Creates an experimental timeline for optical pumping.

    NOTE:
    The AOM is switched off close to, but before, the opening of the first shutter

    WARNING:
    Shutters are reinitialized so that additional optical pumping stages can be added later.
    """

    duration_full = duration_exposition + duration_coil_ramp
    return tl.stack(
        tl.ramp(
            coil__MOT_lower__A=i,
            coil__MOT_upper__A=-i,
            duration=duration_coil_ramp,
        ),
        tl.update(AOM__OP=[[-0.1, 0], [duration_coil_ramp, 1], [duration_full, 0]]),
        tl.update(
            shutter__OP1=[
                [duration_coil_ramp - constants.OP.lag_shutter_on + delay1, 1],
                [delay_shutter_reinitialization, 0],
            ]
        ),
        tl.update(
            shutter__OP2=[
                [duration_full - constants.OP.lag_shutter_off + delay2, 0],
                [delay_shutter_reinitialization, 1],
            ]
        ),
        tl.update(
            shutter__repump=0,
            time=duration_full - constants.lag_repump_shutter + delay_repump,
        ),
        tl.update(AOM__repump=0, time=duration_full),
        tl.anchor(duration_full),
        context="optical_pumping",
    )


def pull_coils(duration, lower_current, upper_current, pt=3, time=None):
    """
    Pull the coil currents to a given value
    """
    return tl.ramp(
        coil__MOT_lower__A=lower_current,
        coil__MOT_upper__A=upper_current,
        function=functools.partial(ramp_function.tanh, sharpness=pt),
        duration=duration,
        time=time,
    )


def magnetic_trapping(
    duration_initial=50e-6,
    lower_current_initial=-1.8,
    upper_current_initial=-1.7,
    duration_strengthen=3e-3,
    lower_current_strengthen=-4.8,
    upper_current_strengthen=-4.7,
):
    """
    Pulls up the magnetic trap
    """
    return tl.stack(
        pull_coils(duration_initial, lower_current_initial, upper_current_initial),
        pull_coils(duration_strengthen, lower_current_strengthen, upper_current_strengthen, time=duration_initial),
        tl.anchor(duration_initial + duration_strengthen),
        context="magnetic_trapping",
    )


###########################################################################
#                   Stage composition                                     #
###########################################################################

timeline_demo = tl.to_timeline(tl.cascade(
    init,
    MOT,
    MOT_detuned_growth,
    molasses,
    optical_pumping,
    magnetic_trapping,
    finish,
    # KW args forwarded to above functions
    init_MOT_ON=True,
    finish_MOT_ON=True,
    MOT_duration=15,
    MOT_lower_current=-1.0,
    MOT_upper_current=-0.98,
    MOT_detuned_growth_duration=0.1,
    MOT_detuned_growth_duration_ramp=1e-2,
    MOT_detuned_growth_detuning__MHz=-5,
    molasses_duration=4.5e-3,
    molasses_duration_coil_ramp=9e-4,
    molasses_duration_lockbox_ramp=1e-3,
    molasses_to__MHz=-90,
    molasses_delay=-200e-6,
    optical_pumping_duration_exposition=80e-6,
    optical_pumping_duration_coil_ramp=500e-6,
    optical_pumping_i=-0.12,
    optical_pumping_delay1=-350e-6,
    optical_pumping_delay2=450e-6,
    optical_pumping_delay_repump=0,
    magnetic_trapping_duration_initial=50e-6,
    magnetic_trapping_lower_current_initial=-1.8,
    magnetic_trapping_upper_current_initial=-1.7,
    magnetic_trapping_duration_strengthen=3e-3,
    magnetic_trapping_lower_current_strengthen=-4.8,
    magnetic_trapping_upper_current_strengthen=-4.7,
))

###########################################################################
#                   Running the experiment
###########################################################################

wtfile.save(timeline_demo)
machine = adwin.link_device()
log = adwin.upload(timeline_demo, connections, devices, machine, process=1) # assuming that the ADwin backend program is loaded as process 1
adwin.run(log) # starts the run, and returns once it has ended
\end{minted}

\section{Extending the default state from a downstream module}
\label{sec:forwarding}

A recurring situation in the operation layer is that a module does not describe a new experiment, but \emph{extends} an existing one with additional hardware. In our laboratory a diagnostics module adds absorption-imaging equipment – an imaging shutter, an imaging AOM and a camera trigger – to the base cold-atom experiment of \cref{sec:demonstration}. These three channels take no part in preparing the sample, but they are nonetheless part of the apparatus, and like every other channel they require a defined state at both ends of a run (\cref{sec:context}).

There are two unsatisfying ways to arrange this. The base \python{default_state} can be edited to mention the imaging hardware, which couples the base experiment to equipment it never uses and makes that one function the place where every future module must intervene. Or the diagnostics module can declare a default state of its own, which duplicates the base specification and reintroduces precisely the divergence that having a single \python{default_state} exists to prevent (\cref{sec:discussion}).

Keyword forwarding provides a third way. \python{default_state} accepts \python{**kwargs} and passes them on to its \python{update} call, which reads any keyword it does not itself recognize as a variable and its value (\cref{sec:functions}). The two stages that wrap it, \python{init} and \python{finish}, forward \python{**kwargs} in the same way, so a keyword that neither consumes falls through to the default state and initializes a variable at both ends of the run. Only these three functions accept arbitrary keywords; ordinary stages name their arguments explicitly, since there a misspelled argument would otherwise become a new variable (\cref{sec:stacking}).

\begin{minted}[fontsize=\tiny, breaklines, breakanywhere, frame=none]{python}
# experiment.py - the base apparatus
def default_state(MOT_ON=True, cavity_transmission=True, **kwargs):
    return tl.update(
        lockbox__MOT__MHz=0.0,
        AOM__MOT=1,
        AOM__repump=1,
        shutter__MOT=MOT_ON,
        shutter__repump=MOT_ON,
        shutter__science=cavity_transmission,
        trigger__TC__V=0.0,
        # ... the remainder of the base state
        **kwargs,
    )

def init(MOT_ON=False, cavity_transmission=False, **kwargs):
    return default_state(
        time=-math.inf, context="ADwin_LowInit",
        MOT_ON=MOT_ON, cavity_transmission=cavity_transmission,
        **kwargs,
    )

# finish(**kwargs) forwards to default_state in an analogous way (appendix B)

# diagnostics.py - a module that adds absorption imaging
connections = pd.concat([
    experiment.connections,
    adcon.new(["shutter__imaging", 1, 13],
              ["AOM__imaging", 1, 5],
              ["trigger__camera", 1, 0]),
])

def init(**kwargs):
    return experiment.init(shutter__imaging=0, AOM__imaging=1, trigger__camera=0, **kwargs)

def finish(**kwargs):
    return experiment.finish(shutter__imaging=0, AOM__imaging=1, trigger__camera=0, **kwargs)
\end{minted}

The three keywords given in \python{diagnostics.init} are consumed by nothing on the way down: \python{experiment.init} passes them on, and \python{default_state} adds them to its \python{update} call. The base experiment file is untouched, the base state is stated exactly once, and \python{diagnostics.finish} does the same against \python{experiment.finish}, so that the added channels are defined at the end of a run as well as at its beginning. The connection table is extended in the same spirit – the diagnostics module declares only its own three channels and concatenates them onto the ones it inherits. This is the module referred to as \python{di} in \cref{sec:parameter_scan}.

Nothing here is a library feature. It is ordinary Python keyword forwarding, available because stages are ordinary functions rather than objects with fixed interfaces, and because \python{update} and \python{ramp} deliberately read any keyword they do not recognize as a variable. The cost of that openness is that a misspelled keyword reaching the final \python{update} call is still a legal variable name, and so creates a variable rather than raising an error. The connection table is what bounds the consequence – a variable with no matching connection cannot reach the hardware, and is removed during conversion (\cref{sec:hardwarelayer}) – which is a further reason to treat that table, rather than the stage functions, as the authoritative statement of what the apparatus is.

\section{Parameter scanning}
\label{sec:parameter_scan}
The functional approach extends naturally beyond single-shot timeline construction into experimental automation. The following example illustrates a time-of-flight (TOF) measurement, in which the same base timeline is composed with an imaging stage at a series of different delay times. The key observation is that this is not specific to TOF: the same pattern applies to any experiment in which a single parameter is scanned while the rest of the timeline remains fixed.
 
The first function is a generator that lazily produces \python{(delay, timeline)} pairs:
\begin{minted}[fontsize=\footnotesize, frame=none]{python}
def tof_timelines(delays, exposure, base_timeline, origin,
                  imaging_function=di.imaging_absorption, **kwargs):
    for d in delays:
        yield d, tl.to_timeline(
            imaging_function(1e-3 * d, exposure, origin=origin, **kwargs),
            onto=base_timeline,
        )
\end{minted}
Since \python{tof_timelines} is a generator, no timelines are constructed until they are consumed. The base timeline, which can be expensive to build, is built once and passed in as a table, and each imaging stage is placed onto it with \python{onto} (\cref{sec:to_timeline}). Note also the role of the \python{origin} argument: passed directly to the imaging function, it determines the reference point within \python{base_timeline} from which the imaging delay is measured. This is the interweaving mechanism of \cref{sec:interweaving} in action – the imaging stage can be hooked into any anchor or context of the base timeline, giving considerable flexibility without modifying the base timeline itself. The second function performs the actual hardware execution. For each timeline, \python{adwin.upload} writes it to the controller and returns a record of the upload, which names the machine and the process to be started. It can therefore be placed directly inside a dict comprehension, yielding a concise one-statement implementation:
\begin{minted}[fontsize=\footnotesize, frame=none]{python}
def parameter_scan_with_imaging(machine, exposure, iterator, **kwargs):
    return {parameter: cc.take_images_ueye(
                4, (exposure + 2 * di.camera_exposition_margin),
                adwin.upload(timeline, connections, devices, machine, process=1),
                **kwargs
            ) for parameter, timeline in iterator}
\end{minted}
This separation – between pure timeline generation and impure hardware execution – is a direct consequence of the functional design of Wigner Time. The timeline construction remains side-effect free and testable independently of any hardware, while the execution loop is kept minimal and generic. The \python{parameter_scan_with_imaging} function is entirely agnostic about what the iterator produces: it could equally well scan a magnetic field, a laser frequency, or any other experimental parameter, as long as the iterator yields \python{(parameter, timeline)} pairs.
 
This is also the point at which the extension to closed-loop operation, noted in \cref{sec:discussion}, becomes concrete: an optimizer that consumes the analysis of one shot and proposes the parameters of the next enters here, as nothing more than a different source for the iterator.


\section{The ADwin back-end in operation}
\label{sec:adwin_operation}
Besides the sequencer – the real-time program of \cref{sec:adwin} – the controller runs a manual console, with which channels are set by hand during alignment and debugging. The two share the controller’s outputs, and this appendix describes how they divide them.

The sequencer checks what it is given before it plays it. \python{upload} records the cycle period at which it converted the timeline, and the sequencer compares this with its own before the first cycle. If the two differ, it plays the initial state and then the final state, and nothing in between, since every time in the experiment would otherwise be rescaled. The rows of the final state, those in the \python{"ADwin_Finish"} context, are applied however the run ends: when it completes, and when it is stopped by hand.

The console is driven by the same \python{connection} and \python{device} tables as the timelines. A channel is therefore named, calibrated and bounded in one place, and a value set on the console reaches its DAC as the same value in a timeline would, through a calibration file where there is one. For each channel the console keeps the code it has been asked for and the code it has last written, and writes the channel only when the two differ, so setting a value never waits for the controller.

The outputs belong to one program at a time. Starting a sequence stops the console, and the console is started again once the sequence has applied its final state. Values set on the console during a run are discarded: after the run the apparatus holds its final state, and the console takes this over as its own, marking as unknown any channel the final state does not name. Before a run, \python{upload} warns about each analog channel that has been set by hand and that the run will change in a single step, whether in the initial state or when a stage first sets it. In our experiment the MOT coils are the case in point. The initial state leaves them untouched, so that the MOT stays in its steady state between runs, and after they have been adjusted by hand the MOT stage would otherwise move them without comment.

\end{appendix}


 
 
 
\bibliography{WignerTime.bib}
 

\end{document}